\documentclass[journal]{IEEEtran}

\usepackage{cite}
\usepackage[pdftex]{graphicx}
\usepackage{subcaption}
\usepackage[font=small]{caption}
\usepackage{textcomp}
\usepackage{gensymb}
\graphicspath{{Figures/}}
\usepackage{amsmath}
\usepackage{amssymb}
\DeclareMathOperator*{\argmin}{argmin}
\DeclareMathOperator*{\argmax}{argmax}

\usepackage{mathrsfs}
\usepackage{mathtools}
\usepackage{algorithmicx}
\usepackage{multirow}
\usepackage[ruled, vlined, linesnumbered]{algorithm2e}
\usepackage{algpseudocode}
\usepackage{booktabs}
\usepackage{array}
\usepackage{color,soul}
\usepackage{xcolor}
\usepackage{mathtools}
\usepackage[nocomma]{optidef}

\begin{document}

\author
{Pengyi~Jia,~\IEEEmembership{Member,~IEEE,}
       Xianbin~Wang,~\IEEEmembership{Fellow,~IEEE,}
       and Xuemin~(Sherman)~Shen,~\IEEEmembership{Fellow,~IEEE}

\thanks{Pengyi Jia and Xianbin Wang are with the Department of Electrical and Computer Engineering, Western University, London, ON N6A 5B9, Canada (e-mail: pjia7@uwo.ca, xianbin.wang@uwo.ca).}
\thanks{Xuemin (Sherman) Shen is with the Department of Electrical and Computer Engineering, University of Waterloo, Waterloo, ON N2L 3G1, Canada (e-mail: sshen@uwaterloo.ca).}
\thanks{(Corresponding author: Xianbin Wang.)}

}
\title{Hierarchical Digital Twin for Efficient 6G Network Orchestration via Adaptive Attribute Selection\\ and Scalable Network Modeling}

%\title{Hierarchical Digital Twin with Attribute Differentiation and Virtual-Real \\ Synchronization for 6G Network Orchestration}

%\title{Hierarchical Digital Twin Enabled 6G Network Orchestration with Attribute Differentiation and Virtual-Real Synchronization}

\maketitle
\begin{abstract}

%The dramatically increased service diversity and infrastructure complexities in forthcoming sixth-generation (6G) networks present formidable challenges in efficient and real-time resource orchestration. 

%Overcoming this issue requires accurate and timely insights into network dynamics. 

%Obtaining a holistic and long-term understanding through network modeling is fundamental in orchestrating future networks with increasing service diversity and infrastructure complexities.

%through network modeling is essential to support rapid network situation evaluation and identification of actual needs and potential issues within the complex network to facilitate targeted and timely decisions for efficient QoS satisfaction. 

Achieving a holistic and long-term understanding through accurate network modeling is essential for orchestrating future networks with increasing service diversity and infrastructure complexities. However, due to unselective data collection and uniform processing, traditional modeling approaches undermine the efficacy and timeliness of network orchestration. Additionally, temporal disparities arising from various modeling delays further impair the centralized decision-making with distributed models. In this paper, we propose a new hierarchical digital twin paradigm adapting to real-time network situations for problem-centered model construction. Specifically, we introduce an adaptive attribute selection mechanism that evaluates the distinct modeling values of diverse network attributes, considering their relevance to current network scenarios and inherent modeling complexity. By prioritizing critical attributes at higher layers, an efficient evaluation of network situations is achieved to identify target areas. Subsequently, scalable network modeling facilitates the inclusion of all identified elements at the lower layers, where more fine-grained digital twins are developed to generate targeted solutions for user association and power allocation. Furthermore, virtual-physical domain synchronization is implemented to maintain accurate temporal alignment between the digital twins and their physical counterparts, spanning from the construction to the utilization of the proposed paradigm. Extensive simulations validate the proposed approach, demonstrating its effectiveness in efficiently identifying pressing issues and delivering network orchestration solutions in complex 6G HetNets.
\end{abstract}


\section{Introduction}
\IEEEPARstart{A}{s} we progress toward the advent of sixth-generation (6G) telecommunications, a transformative shift in network architecture is witnessed, marked by tremendously increased heterogeneity and complexity \cite{Roadto6G}. Anticipated 6G heterogeneous networks (HetNets) are distinguished not only by the proliferation of base station (BS) types, including macro, micro, and femto BSs, but also through the diversification in quality of service (QoS) requirements to support various vertical applications, such as Extended Reality, Metaverse, and Internet of Everything \cite{Realizing6G}. Amidst this expeditious advancement, the challenge of managing burgeoning data traffic associated with the intricate interactions among wireless devices with constrained spectrum resources becomes increasingly formidable, thereby highlighting the pressing necessity for efficient QoS requirement satisfaction \cite{JSAC}.

\subsection{Motivations}
Given the heterogeneity and complexity of 6G networks, it is necessary to design innovative network orchestration strategies, including resource allocation and traffic engineering, to efficiently satisfy the heterogeneous QoS requirements within complex network architectures \cite{HetNetsSurvey}. %1, 6G HetNet的需求
Current network orchestration is typically achieved through network-wide large-scale optimization, such as network congestion control \cite{CC1} and BS-user association \cite{CC2}, which rely on comprehensive data acquisition across the network and centralized processing for managing all involved infrastructures.  %2， 如今方法的实现方式

However, with the surge in communication devices and their associated diverse network attributes in 6G HetNets, collecting and processing data for every attribute from each user to support large-scale network optimization places an excessive burden on the processing center \cite{MLforOP}. %3，随着网络规模的变大，如今方法的不足。 
Due to limited computational capability and spectrum resources, existing aimless and unstructured orchestration methods could inevitably result in suboptimal performance and delayed decision-making in highly dynamic 6G HetNets. %4，动态
This necessitates a more scalable and adaptive orchestration strategy, rather than existing rigid, one-size-fits-all solutions, in efficiently satisfying user QoS demands under various network situations. %1，以上方法的不足指出需要一种可扩展的自适应方法以面向网络动态
In light of this, attaining a holistic and long-term insight into network situational dynamics through network modeling becomes essential, which allows rapid identification of actual needs and potential issues within the complex network to facilitate targeted and timely decisions for efficient QoS satisfaction. %2，基于此，对网络的动态有长期而全面的理解变得极其重要。 %3，这种理解可以帮助快速的界定网络状况，指出网络问题，并产生准确及时的决策以支持网络管理。

To understand network situations and pinpoint pressing issues, digital twin technology offers a promising solution for real-time and comprehensive modeling of communication networks in remote control centers \cite{DT2, DTTCOM}. By harnessing statistical data on network performance indicators, digital twins facilitate the rapid analysis of large-scale networks, circumventing the reliance on the highly dynamic physical layer attributes \cite{JSAC}. The instantaneous and statistical link information generated from digital twins can be instrumental in informing decisions across various layers of the communication systems \cite{RealTimeDT, syncSurvey}. Therefore, digital twins emerge as an indispensable tool for achieving efficient orchestration and pervasive intelligence within 6G networks, enabling timely problem identification and bolstering decision-making in complex networks \cite{ShenDT}. 

However, current digital twin-empowered network orchestration methods, which utilize fine-grained simulation of the entire network, encounter the following significant challenges in efficiently evaluating network situations and identifying problems within complex 6G HetNets.

\textbullet~The extreme modeling complexity of creating comprehensive and fine-grained digital twins for large-scale 6G HetNets stems from the myriad of heterogeneous wireless devices, including a massive array of mobile users and various types of opportunistically deployed BSs \cite{HetNet1}. Further complexity is introduced by the dynamic QoS requirements and the fluctuating communication environments of distributed users. These factors combine to produce a wide array of network situations, characterized by varying QoS provisioning, non-uniformly distributed BS loads, and inconsistent network resource availability \cite{HetNet2}. Implementing a uniform approach to digital twin construction and network orchestration in such a dynamic and heterogeneous setting can significantly hinder responsiveness to real-time network issues, particularly for devices in suboptimal conditions, while also risking resource wastage for devices in more stable environments. Therefore, accurately capturing essential network dynamics and effectively prioritizing urgent issues among the complex and intertwined network components becomes crucial for efficient network orchestration.

\textbullet~The oversight of situation-dependent modeling values for heterogeneous network attributes during digital twin construction. The cost of modeling different attributes can vary widely, influenced by factors such as the volume of data required, the inherent complexity of the data, and the stability of its patterns. Additionally, the relevance of these attributes to current user requirements and network demands fluctuates with network dynamics, making the value of each attribute highly situation-specific. For instance, attributes that are computation-intensive due to their dynamic and complex features may still hold considerable value if they directly address the most pressing issues in the current network scenario. Given limited communication resources and the necessity for real-time optimization, this necessitates a transition towards a value-based modeling approach for complex and dynamic 6G HetNets, diverging from traditional methods that unselectively model all network attributes without regard to their immediate utility.

\textbullet~Temporal misalignment between wireless devices and their digital twins due to their intrinsic physical heterogeneity and excessive modeling delay. The inconsistency of local clocks and the variance in sampling capabilities across physical devices inevitably lead to discrepancies in modeling data and obscure temporal correlation \cite{syncMagazine, syncTCOM}. Moreover, the diverse processes involved in digital twin construction, from data transmission to centralized processing, introduce uniquely accumulated delays that exacerbate the temporal misalignment among digital twins \cite{sync4}. As a direct result, the interactions across physical and digital realms will inevitably suffer from significant temporal disparities. This manifests as a misalignment between the outputs of digital twins and the actual real-time network conditions \cite{sync3}. Conventional methods, which overlook these synchronization challenges, will result in insights that lack actionable value, thus undermining their utility in enabling timely and responsive network orchestration.

Based on these observations, it is essential to design a novel digital twin paradigm cohesively considering network modeling complexity, heterogeneous modeling value, and virtual-physical synchronization issues during the design in supporting efficient network orchestration.

\subsection{Related Work}
The profound capabilities of digital twins in mirroring dynamic network states for network orchestration have been studied recently. For example, a graph neural network (GNN) based digital twin framework is proposed in \cite{DTN1} to achieve optimal power allocation and user association in the THz band. By leveraging the prediction from GNN, efficient topology mining and feature extraction are achieved within the digital twin network. Digital twins can be used to simulate and analyze the dynamic characteristics of the THz band for network orchestration. Furthermore, digital twins are utilized for integrated user association, task offloading, and resource allocation in ultra-reliable and low-latency communication (URLLC) networks \cite{DTN2}, focusing on reducing offloading latency. Playing a vital role in enabling informative analysis of edge servers, digital twins could still face challenges due to complex spatial-temporal dynamics in URLLC networks, which hinder the effectiveness of timely decision-making. To tackle this issue, a digital twin mobile network is designed in \cite{JSAC} to utilize the virtual network topology that effectively manages the spatial-temporal dynamics and load distribution in the network. The overall network performance and QoS satisfaction are significantly improved during user association with balanced inclusiveness and selectiveness during digital twin construction so that the complex dynamics of HetNets can be efficiently handled.

On the other hand, cross-domain synchronization is critical for digital twins to address the misalignment between the digital domain and its physical counterpart, thus ensuring the virtual models accurately reflect their physical features in real time \cite{sync0}. Based on this observation, a dynamic framework for synchronizing digital twins in Metaverse is proposed in \cite{sync3}, which focuses on aligning digital twins for virtual service providers with real-world data generated by IoT devices. The strategy involves a game-theoretic approach to optimize the frequency and intensity of synchronization, ensuring that digital twins accurately reflect their physical counterparts. Similarly, by adopting a network selection algorithm based on a learning-based prediction model, accurate data synchronization is achieved between vehicular digital twins and vehicular users in heterogeneous vehicular networks \cite{sync1}. By predicting and compensating for the data transmission waiting time, the latency of digital twins can be reduced with improved overall reliability and responsiveness of the system. However, the temporal misalignment among physical data is not considered. By contrast, authors in \cite{sync4} present a novel method for synchronizing and resampling multi-attribute data in supporting digital twin construction. The temporal misalignment in data due to asynchronous IoT devices and heterogeneous sampling rates is addressed, although the impact of multi-source delays on virtual-physical domain misalignment is overlooked.

Obviously, existing digital twin-based schemes always ignore the complexity and various delays induced by modeling the entire network. The huge communication and computational resources consumed during digital twin construction, alongside the temporal misalignment induced by processing and transmission delay, can significantly degrade the modeling efficiency and accuracy in large-scale and highly dynamic HetNets. Based on this observation, we previously proposed a dual-layered digital twin paradigm in \cite{InfoCom} to identify hotspots in small 6G HetNets for intelligent network orchestration. However, the inefficiency of network situation discovery and the lack of cross-domain synchronization remain unsolved for highly complex and heterogeneous future networks.

\subsection{Contributions}
Motivated by these observations, we propose a hierarchical digital twin paradigm to support efficient network orchestration enabled by adaptive attribute selection and scalable network modeling, with the following main contributions.

\textbullet~Decomposing the complex network modeling into the rapid evaluation of network situations and problem-centered model construction through hierarchical digital twins for efficient network orchestration. In the higher layers, coarse-grained digital twins focus on system-level network evaluation, enabling scalable selection of target areas and critical network attributes vital for enhancing orchestration efficiency. Subsequently, fine-grained lower-layered digital twins for the selected elements are modeled to ensure efficient QoS satisfaction through predictive user association and power allocation strategies. This decomposition facilitates a more prioritized and effective management approach that adaptively satisfies the dynamic service requirements in complex 6G HetNets.

\textbullet~Prioritizing network attributes based on their modeling benefit and associated costs in supporting network orchestration at higher layers. An in-depth analysis of the modeling value is provided for differentiating network attributes according to their relevance to current network situations evaluated by temporal correlation and causality inferences, as well as the inherent modeling complexity as indicated by sample entropy. Such a prioritized approach adapts to real-time network dynamics, allowing for allocating more resources to model highly valued attributes at higher layers for rapid network situation assessment and problem identification. Conversely, lower-valued attributes due to high modeling cost are selectively modeled in lower layers with finer granularity, where additional resources can be allocated for comprehensive analysis. Through attribute differentiation and prioritization, this value-oriented mechanism ensures the efficient inclusion of essential elements in digital twin modeling with dramatically reduced resource wastage.

\textbullet~Implementing virtual-physical domain synchronization for accurate temporal alignment between digital twins and their physical counterparts to support efficacious model utilization in lower layers. By focusing on device-level synchronization across individual devices, temporal disparities caused by clock inaccuracy and diverse sampling capabilities are first addressed. Subsequent model creation utilizes the nonlinear auto-regressive exogenous neural network (NARX), incorporating seasonal and trend decomposition using Loess (STL) to parse network patterns. To counteract misalignment between digital twins, which stems from delays inherent in various modeling processes, compensation is made for the time lag between data collection and model utilization to ensure the real-time reflection of network conditions.

The remainder of this paper is organized as follows. In Section II, the network orchestration problem is formulated for the complex 6G HetNets, where the hierarchical digital twin paradigm is introduced as a potential solution. Higher-layered problem identification and lower-layered problem-centered network modeling are given in Sections III and IV, respectively. In Section V, we demonstrate the simulation results to evaluate the effectiveness of the proposed method, followed by the conclusion in Section VI.
\section{System Model}
\label{systemModel-section}
In this section, we present the problem formulation to maximize the network orchestration efficiency in large-scale 6G HetNets. Given its inherent complexity, a hierarchical digital twin paradigm is then proposed as an effective solution through complex problem decomposition.

\subsection{Heterogeneous Networks Model}
\begin{figure*}[t]
    \centering
    \includegraphics[width=17cm]{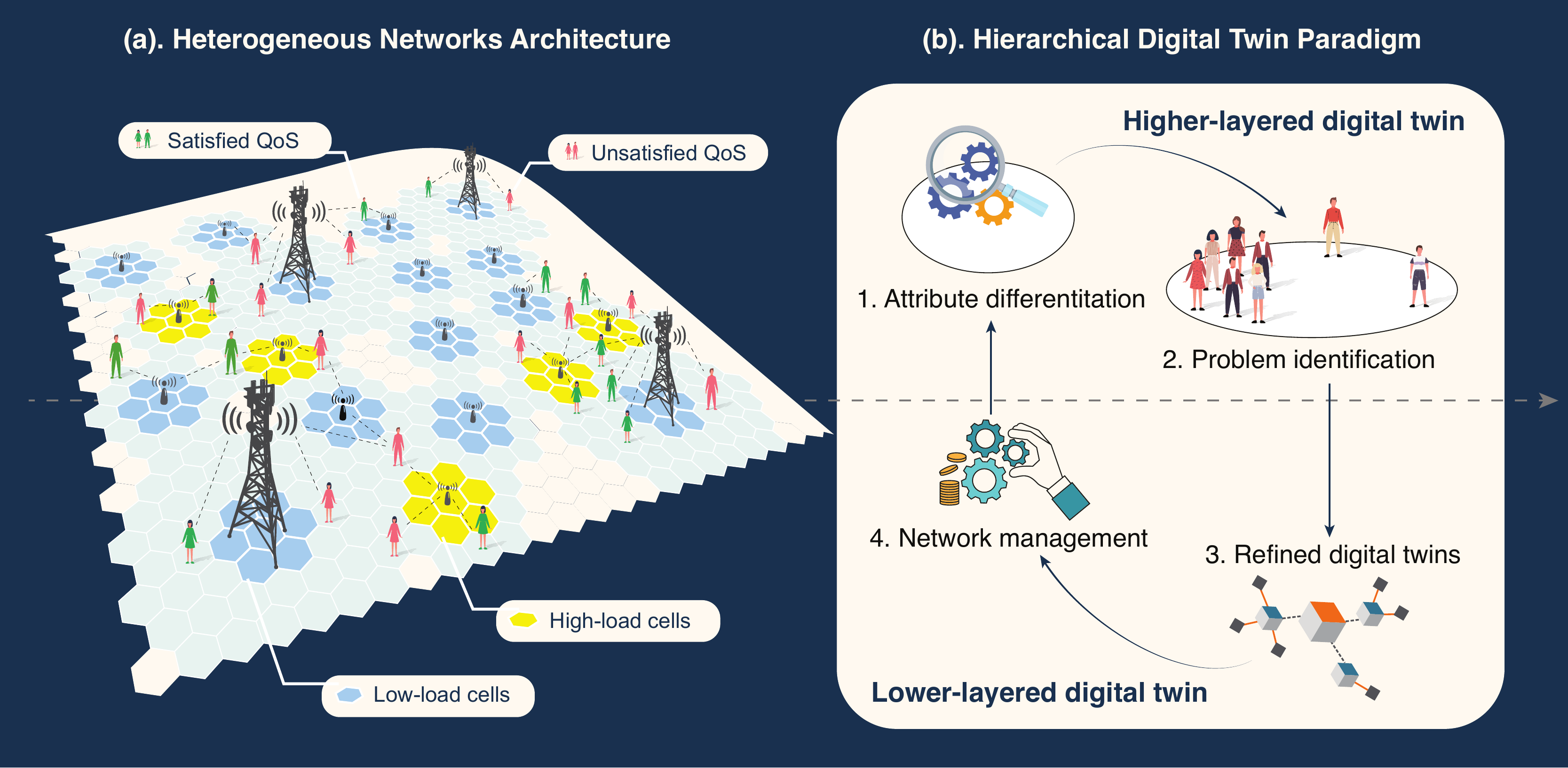}
    \caption{The overall architecture of 6G HetNets orchestrated by the proposed hierarchical digital twin paradigm: a). The overall 6G HetNets, where macro-cell and small-cell BSs serve the distributed users to accommodate their QoS requirements. Unsatisfied QoS is caused by unbalanced traffic load distribution and insufficient network capacity for local cells. b). Hierarchical digital twins for efficient network orchestration, comprising higher layers for network problem identification and lower layers for problem-focused network management.}
    \label{fig:main}
\end{figure*}

Consider 6G HetNets shown in Fig. \ref{fig:main}(a), a total of $M$ BSs $\mathcal{M}=\{1,...,M\}$ with different locations and transmission powers will be deployed to serve $N$ mobile users $\mathcal{N}=\{1,...,N\}$ with diverse QoS requirements. A graphical representation of the network can be given by $G=(\mathcal{V},E)$, with vertices $\mathcal{V}=\{\mathcal{M}, \mathcal{N}\}$ for all infrastructures in the network as well as the edge $E$ for the communication link between each user and its associated BS. 

According to the current user association strategy, each user will be served by a specific BS at a time, while the frequent handover between different BSs will lead to complex traffic throughout the network. Within this complexity, high-loaded BSs and QoS-unsatisfied users due to unoptimized network orchestration are considered the bottleneck of the entire network. To tackle this issue, the data traffic should be properly engineered by adjusting the user association and power allocation strategy according to the network situation to fulfill the stringent QoS requirements for all users. The traffic engineering performance can be evaluated through the satisfaction level of application-related QoS requirements for all users, e.g., network throughput, packet loss rate, and transmission delay. Denoted by $Q_{i,j}^n$, the $n$th QoS fulfillment of user $i$ is highly relevant to the instantaneous downlink data rate it received from the associated BS $j$, given by
\begin{equation}\label{dataRate}
  r_{i,j}  = B\log_2\Bigl(1+\frac{p_{i,j}|h_{i,j}|^2}{\sum_{l\in\mathcal{M},l\neq j}p_{i,l}|h_{i,l}|^2+\sigma_0^2}\Bigr),
\end{equation}
where $B$ and $\sigma_0$ are the bandwidth and power of background noise. The signal strength of user $i$, $p_{i,j}|h_{i,j}|^2$, is determined by the transmission power $p_{i,j}$ from its associated BS $j$ and the Rayleigh flat fading channel coefficient $h_{i,j}$, which is further affected by the path loss model and distance. Moreover, we only consider the inter-cell interference $\sum_{l\in\mathcal{M},l\neq j}p_{i,l}|h_{i,l}|^2$ with universal frequency reuse.

Based on the instantaneous data rate obtained in (\ref{dataRate}), the satisfaction level for different QoS requirements of user $i$ can be evaluated by
\begin{equation}\label{QoSRatio}
  S_{i}^n = Q_{i}^n/{\widehat{Q}_{i}^n},
\end{equation}
where $\widehat{Q}_{i}^n$ is the $n$th QoS requirement associated to user $i$. The $n$th condition $Q_{i}^n = \sum\nolimits_{j\in\mathcal{M}}{Q_{i,j}^n}$ relates to the quality of link $i$-$j$ for each association. QoS conditions can be evaluated based on various network attributes about link quality and network situations. Taking achievable data rate as an example, the corresponding QoS for user $i$ during network orchestration can be written as
\begin{equation}
    \label{QoS1}
    Q_i^{dr} = \sum\nolimits_{j\in\mathcal{M}}a_{i,j}r_{i,j},
\end{equation}
where $a_{i,j}$ is a binary variable indicating the association between user $i$ and BS $j$. Similarly, other QoS metrics, e.g., network delay and packet losses can be derived based on real-time network conditions, indicating the satisfaction of the application requirements. According to the assigned application, different users will have unique QoS metrics, thus requiring a scalable and user-specific network orchestration strategy to ensure QoS satisfaction.

\subsection{Problem Formulation}
In satisfying diverse user QoS demands, selecting the appropriate BS for data transmission with optimal transmission power is critical. Therefore, in this paper, network orchestration will focus on user association factor $a_{i,j}$ and transmission power $p_{i,j}$ for each user to ensure overall satisfied QoS requirements. Meanwhile, real-time network orchestration requires frequent data transmission and processing from distributed users for network optimization, leading to excessive consumption of limited network resources. Typically, for the same achievable QoS satisfaction level, an algorithm with less data transmission will always lead to a faster response to the dynamic network situation and efficient utilization of network resources, thus bringing higher benefits for the entire network. 

Based on this observation, we formulate the problem to maximize the QoS satisfaction efficiency, defined as the QoS satisfaction level over the resource consumed during network orchestration. For clarity, we define the user association matrix as $\mathbf{a}=\{a_{i,j} \vert\ \forall i\in\mathcal{N}, \forall j\in\mathcal{M}\}$ and power allocation matrix as $\mathbf{p}=\{p_{i,j} \vert\ \forall i\in\mathcal{N}, \forall j\in\mathcal{M}\}$, respectively. The problem of network orchestration can be written as
\begin{align}
%\label{Solution0}
{\bf {\cal P}1}:\mathop {\max }\limits_{\mathbf{a}, \mathbf{p}}\;\;&{\cal E}_Q\left( {\mathbf{a}, \mathbf{p}} \right)= \frac{\sum_{i\in \mathcal{N}}\sum_{j\in\mathcal{M}}\sum_{n\in\mathcal{Q}}a_{i,j} \omega_{i}^nS_{i}^n(\bf p)}{\sum_{i\in\mathcal{N}}\sum_{n\in\mathcal{Q}}\boldsymbol{\alpha}\mathcal{R}_{i}^n}\nonumber \\
\quad{\rm {s.t.}}\quad&{\rm C1:}\; a_{i,j} \in \{0,1\}, \quad \forall i\in \mathcal{N}, \forall j\in \mathcal{M}, \nonumber \\
&{\rm C2:}\; \sum\nolimits_{j\in \mathcal{M}}a_{i,j}\leq 1,\quad\forall i\in \mathcal{N}, \nonumber \\
&{\rm C3:}\; {\sum\nolimits_{i\in \mathcal{N}}a_{i,j}}{\leq 1,}\quad{\forall j\in \mathcal{M}}, \nonumber \\
&{\rm C4:}\; {p_{i,j}}{\geq0,}{\quad\forall i\in \mathcal{N},\forall j\in \mathcal{M}}, \nonumber \\
&{\rm C5:}\; {\sum\nolimits_{i\in \mathcal{N}}a_{i,j}p_{i,j}}{\leq P_{j,\max},}{\quad\forall j\in \mathcal{M}}, \nonumber \\
&{\rm C6:}\; {S_{i}^n}{\geq1,}{\quad\forall i\in \mathcal{N},\forall n\in \mathcal{Q}}, \nonumber
\end{align}
where the set $\mathcal{Q}=\{1,...,N_Q\}$ is a collection of all QoS metrics weighted by $\omega_{i}^n$ according to the application preference. The scaling vector $\boldsymbol{\alpha}$ will be a large number for users with satisfied QoS requirements, thus ensuring prioritized resource allocation for users with $S_{i}^n<1$. ${\rm C1}$ to ${\rm C3}$ are the constraints of the user association basis. ${\rm C4}$ and ${\rm C5}$ are related to the transmission power of each BS, limited by its maximized power $P_{j,\text{max}}$. Moreover, ${\rm C6}$ ensures the QoS satisfaction for all users during network orchestration.

Furthermore, $\mathcal{R}_{i}^n$ is the network resource consumed to transmit the network attribute data related to user $i$ regarding the $n$th QoS metrics. By adopting different optimization methods, e.g., integer relaxation and problem decomposition \cite{MixedIntegar}, the complex network orchestration problem could be solved. However, the overall resource consumption due to the need for all-inclusive collection of current network data from all users and attributes will increase exponentially to orchestrate the entire network. In contrast, the improvement of QoS satisfaction for different users and QoS metrics will not be identical throughout the network. As a result, the current network orchestration strategy that treats all users and attributes equally will significantly degrade the real-time network orchestration efficiency for large-scale 6G HetNets.

\subsection{Digital Twin-Assisted Predictive Assignment Problem}
Substantially, $\bf {\cal P}1$ is a nonlinear mixed integer programming problem that is complex to solve. For clarity, we consider $\bf {\cal P}1$ as an assignment problem between $N$ users and $M$ BSs. Each edge $i$-$j$ represents one agent-task assignment between user $i$ and BS $j$ with achievable QoS $Q_{i}^n$ under the current network orchestration strategy \{$\mathbf{a}$, $\mathbf{p}$\}. In other words, different assignment strategies will lead to various assignment awards $A_{i,j} = \sum_{n\in\mathcal{Q}}\omega_{i}^nS_{i}^n(\bf p)$ and assignment costs $c_{i}=\sum_{n\in\mathcal{Q}}\boldsymbol{\alpha}\mathcal{R}_{i}^n$, so that $\bf {\cal P}1$ can be rewritten as
\begin{align}
\label{Eq:4}
{\bf {\cal P}2}:\quad \mathop {\max }\limits_{\mathbf{a}, \mathbf{p}}\;\;&{\cal E}_Q\left( {\mathbf{a}, \mathbf{p}} \right)= \frac{\sum\nolimits_{(i,j)\in \mathcal{N}\times \mathcal{M}}a_{i,j}A_{i,j}}{\sum_{i\in\mathcal{N}}c_i} \\
\quad{\rm {s.t.}}\quad&{\rm C1 - \rm C6}.\nonumber
\end{align}

However, due to the combination of binary variable $\mathbf{a}$ and continuous variable $\mathbf{p}$ as well as the joint consideration of assignment reward $A_{i,j}$ and cost, the discrete and nonlinear nature of $\bf {\cal P}2$ significantly limits the available solutions. Existing studies mainly rely on heuristic algorithms or reinforcement learning \cite{AssignDRL} \cite{TE1}, which are likely to induce significant processing complexity with sub-optimal solutions, thus degrading the timeliness and effectiveness of decision-making for real-time network orchestration. 

By contrast, a digital twin-assisted linear mixed-integer programming method was discussed in \cite{JSAC} to maximize QoS satisfaction through user association, where statistical network attributes are modeled to predict the assignment award for different association strategies according to historical user information. The use of prediction can dramatically reduce the need for real-time network data upload, thus improving the network orchestration timeliness and efficiency. Nevertheless, with the increment of network scales and the increasing number of network attributes, the construction of network digital twins could overwhelmingly consume limited network resources and become unaffordable for network orchestration.

Motivated by this observation, the nonlinear problem $\bf {\cal P}2$ can be considered as maximizing the network-wide digital twin-assisted network orchestration efficiency, with award of using digital twins $\Phi(\mathbf{a},\mathbf{p}) = \sum\nolimits_{(i,j)\in \mathcal{N}\times \mathcal{M}}a_{i,j}A_{i,j}$ and corresponding construction cost $\Theta({\cal N},{\cal K}) = \sum_{i\in\mathcal{N}}\sum_{k\in\mathcal{K}}\mathcal{C}_{i,k}$, where $\mathcal{C}_{i,k}$ is the cost for digital twinning the $k$th attribute in user $i$, while $\mathcal{K}$ is the collection of all network attributes. However, the relationship between $\Phi$ and $\Theta$ is not intuitive, where $\Phi$ depends on the various $\{\bf a, p\}$ scenarios, while $\Theta$ is determined by the orchestration target. Therefore, it motivates us to identify the \textit{target areas} $\boldsymbol{\cal Z}$ in the network leading to maximized benefits, instead of selecting all of them. By limiting the areas to be modeled, the overall orchestration efficiency can be substantially enhanced.

%It can be observed that $\Theta$ is independent of $\mathbf{a}$ and $\mathbf{p}$, but instead, determined by the target of network orchestration. By limiting the entities required to be modeled, the overall orchestration efficiency can be substantially enhanced.

\begin{figure*}[t]
    \centering
    \includegraphics[width=14.5cm]{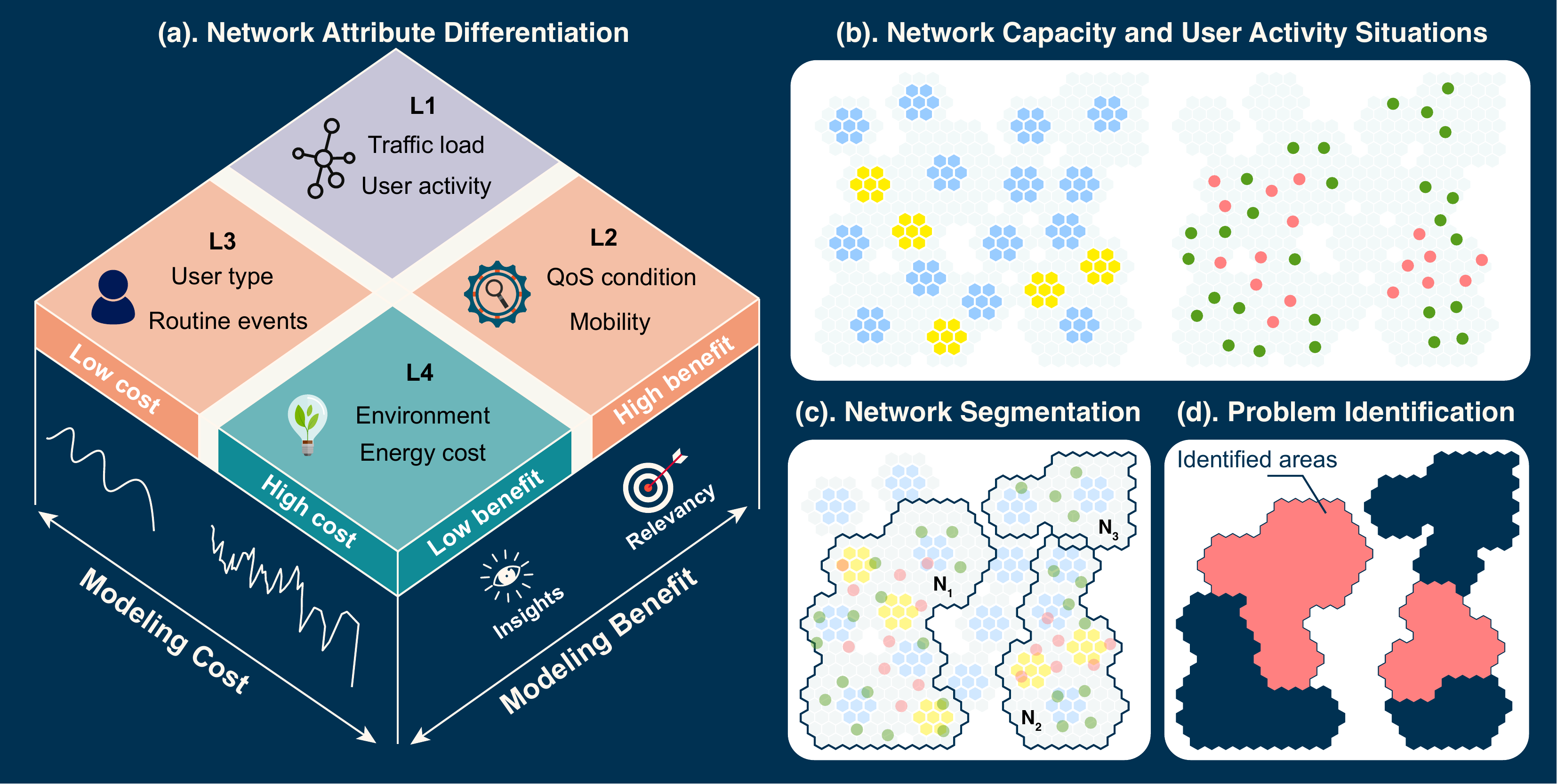}
    \caption{Higher-layered digital twins for network situation evaluation: a). Differentiation and selection of various network attributes based on the cost and benefit of constructing digital twins. b). The overall situation in terms of the network capacity from the perspectives of BSs and user activities. c). Network segmentation according to the historical user activity and potential connections. d). Target areas in each network segment based on the expected network orchestration efficiency.}
    \label{fig:HLDT}
\end{figure*}
\subsection{Hierarchical Digital Twin for Efficient Orchestration}
To identify target areas $\boldsymbol{\cal Z}$ and achieve efficient network orchestration, we propose a novel hierarchical digital twin paradigm, as shown in Fig. \ref{fig:main}(b), to decompose and solve $\bf {\cal P}2$ by analyzing and predicting time-series data from the network. The following two layers are established in the cloud center with different granularity and functions.

\textbullet~\textbf{\textit{Higher-layered digital twins} (HDT)} leverage system-level situation evaluation and network attribute differentiation to rapidly identify target areas $\boldsymbol{\cal Z}$ in the network. Instead of creating detailed models, HDT focuses on constructing simplified, coarse-grained digital twins for highly valued attributes across the system for complexity-reduced modeling. Real-time data from other attributes are selectively utilized to efficiently evaluate network situations and identify problematic users, forming the target areas for further analysis.

\textbullet~ \textbf{\textit{Lower-layered digital twins} (LDT)} establish more fine-grained digital twin models for the target areas $\boldsymbol{\cal Z}$ and attributes identified in HDT. To achieve detailed network orchestration, more advanced data processing and modeling techniques are utilized, thereby enhancing digital twin granularity. Furthermore, LDT employs higher frequency synchronization to ensure precise alignment between the virtual and physical domains. By narrowing the focus to selected elements, LDT significantly enhances orchestration efficiency, offering a problem-centered network management approach.

More specifically, the diversity in user activity patterns and application demands will cause various benefits and costs of establishing digital twins for different users and network attributes. Therefore, to identify target areas with users and attributes leading to the most efficient network orchestration outcome, the objective function of HDT can be written as
\begin{align}
{\bf {\cal P}3}:\quad \mathop {\min }\limits_{\cal N, \cal K}\;\;&{\Theta}\left( {\cal N, \cal K} \right).
\end{align}
Define $\mathcal{N}_s = \{1,\cdots,N_s\}$ and $\mathcal{K}_s = \{1,\cdots,K_s\}$ as the set of users and attributes selected, i.e., $\{\mathcal{N}_s,\mathcal{K}_s\} = \argmin_{\mathcal{N},\mathcal{K}}{\Theta}\left( {\cal N, \cal K} \right)$, the modeling cost can be given by
\begin{equation}
    \widehat{\Theta} = \sum\nolimits_{i\in\mathcal{N}_s}\sum\nolimits_{k\in\mathcal{K}_s}\mathcal{C}_{i,k}.
\end{equation}
It is worth noting that this selection is strictly constrained by the QoS satisfaction level and resource availability in the selected areas, which are specified in Section \ref{HDT-Section}.

Then, LDT will focus on maximizing the QoS satisfaction level for selected users ${\cal N}_s$ based on attributes ${\cal K}_s$, given by
\begin{align}
\label{Solution1}
{\bf {\cal P}4}:\quad \mathop {\max }\limits_{\mathbf{a}, \mathbf{p}}\;\;&\mathcal{E}_Q(\mathbf{a}, \mathbf{p})= \frac{1}{\widehat{\Theta}}\sum\nolimits_{(i,j)\in \mathcal{N}_s\times \mathcal{M}}a_{i,j}A_{i,j}\\
\quad{\rm {s.t.}}\quad&{\rm C1:}\; a_{i,j} \in \{0,1\}, \quad \forall i\in \mathcal{N}_s, \forall j\in \mathcal{M}, \nonumber \\
&{\rm C2:}\; \sum\nolimits_{j\in \mathcal{M}}a_{i,j}\leq 1,\quad\forall i\in \mathcal{N}_s, \nonumber \\
&{\rm C3:}\; {\sum\nolimits_{i\in \mathcal{N}_s}a_{i,j}}{\leq 1,}\quad{\forall j\in \mathcal{M}}, \nonumber \\
&{\rm C4:}\; {p_{i,j}}{\geq0,}{\quad\forall i\in \mathcal{N}_s,\forall j\in \mathcal{M}}, \nonumber \\
&{\rm C5:}\; {\sum\nolimits_{i\in \mathcal{N}_s}a_{i,j}p_{i,j}}{\leq P_{j,\max},}{\quad\forall j\in \mathcal{M}}, \nonumber \\
&{\rm C6:}\; {S_{i}^n}{\geq1,}{\quad\forall i\in \mathcal{N}_s,\forall k\in \mathcal{Q}}, \nonumber
\end{align}
which is a simplified assignment problem since the minimum modeling cost $\widehat{\Theta}$ can be considered a constant for LDT. The accurate solution to $\bf {\cal P}4$ exclusively hinges on the fine-grained modeling of the involved network attributes $\mathcal{K}_s$ for each user, thus necessitating more advanced data processing and modeling techniques as illustrated in Section \ref{LDT-Section}.

\textbf{\textit{Remark:}} The network orchestration problem $\bf {\cal P}2$ prioritizes costs influenced by the scaling vector $\boldsymbol{\alpha}$, which regulates the consumption of additional resources for users with satisfied QoS requirements. Consequently, an optimal solution for $\bf {\cal P}2$ can be equivalently achieved by initially focusing on identifying unsatisfied users and critical attributes through problem $\bf {\cal P}3$, followed by addressing the QoS maximization problem $\bf {\cal P}4$ for these identified elements.

\section{Problem Identification in Higher Layers With Adaptive Attribute Selection}
\label{HDT-Section}

The essential role of HDT in identifying $\mathcal{N}_s$ and $\mathcal{K}_s$ among the massive users and heterogeneous attributes within complex 6G HetNets can be achieved by solving $\bf {\cal P}3$. To accommodate the dynamic nature of network situations and user activities during the identification, we develop an adaptive attribute selection algorithm, which enhances the efficiency of network evaluation and problem discovery, as shown in Fig. \ref{fig:HLDT}.

\subsection{Value-Oriented Network Attribution Selection}
Given the inherent heterogeneity and diversity of network attributes, modeling each one during digital twin construction is inefficient and impractical due to the varying values of these attributes to the assigned application. Therefore, we propose a value-oriented attribute differentiation mechanism to select the most pertinent attributes $\mathcal{K}_s$ within the network according to its actual demands. Moreover, since HDT is primarily used for rapid problem identification, simple yet effective methods will be preferred to enhance the efficiency and response speed.
%as illustrated in Algorithm \ref{NAD-Alg},

As depicted in Fig. \ref{fig:HLDT}(a), the value of each network attribute is evaluated by its modeling cost and benefit. Specifically, the cost of modeling any attribute depends on the predictability, which can be indicated by the complexity of the data. For a series of $X_M$ data $\mathbf{x_{i,k}}=\{x_{i,k}^l\,|\,l=1,...,X_M\}$ used to model the $k$th attribute from user $i$, its complexity can be evaluated via sample entropy \cite{Entropy}, given by
\begin{equation}\label{SampleEntropy}
\text{SampEn}_{i,k} = -\ln\left(\frac{\Psi^{m+1}_\gamma(\mathbf{x_{i,k}})}{\Psi^m_\gamma(\mathbf{x_{i,k}})}\right),
\end{equation}
where $\Psi^m_\gamma(\cdot)$ denotes the conditional probability that a sequence of $m$ consecutive data $\mathbf{x_{i,k}^{m,u}}=\{x_{i,k}^l\,|\,l=u,...,u+m-1\}$ in $\mathbf{x_{i,k}}$ will share similarity within a predefined tolerance level $\gamma$, given by
\begin{equation}
\label{PSI}
    \Psi^m_\gamma(\mathbf{x_{i,k}}) = \frac{1}{X_N-m}\sum\limits_{u=1}^{X_N-m}\sum\limits_{v=1,v\neq u}^{X_N-m}\mathbf{1}_{d(\mathbf{x_{i,k}^{m,u}},\mathbf{x_{i,k}^{m,v}})<\gamma},
\end{equation}
where $\bf 1$ is a binary indicator counting the number of sequences in $\mathbf{x_{i,k}}$ to be similar to $\mathbf{x_{i,k}^{m,u}}$, while $d(\cdot)$ is the Chebyshev distance, showing the maximum absolute difference between elements in two data sequences, written as
\begin{equation}
\label{distance}
d(\mathbf{x_{i,k}^{m,u}},\mathbf{x_{i,k}^{m,v}})=\max\nolimits_l(|x_{i,k}^{m,u}(l)-x_{i,k}^{m,v}(l)|).
\end{equation}
With more sequences of data showing similarity in the $k$th attribute, i.e., $d<\gamma$, a smaller sample entropy $\text{SampEn}$ will be expected, indicating less complex data and lower modeling cost for the $k$th attribute.

%\begin{algorithm}[t]
%\DontPrintSemicolon
%\textbf{Attribute value estimation:} \\
%\For{each user $i\in \mathcal{N}$}
%{
%    \For{each attribute $k\in \mathcal{K}$}
%    {
%        Collect data samples $\mathbf{x_{i,k}}$ \\
%        Obtain a sequence of $m$ data $\mathbf{x_{i,k}^{m,p}}$ from $\mathbf{x_{i,k}}$ \\
%        Calculate modeling cost $\text{SampEn}_{i,k}$ according to Eq. (\ref{SampleEntropy})~--~(\ref{distance}) \\
%        Collect objective data samples $\mathbf{Q_{i}^n}$\\
%        Calculate modeling benefit $\rho_{i,k}^n$ by Eq. (\ref{CorrelationCoefficient}) \\
%        \If{$|\rho_{i,k}^n|<0.3$}
%        {
%            Test Granger causality for $\mathbf{Q_{i}^n}$ and $\mathbf{x_{i,k}}$ \\
%            Update $|\rho_{i,k}^n|$ based on causality \\
%        }
%        Record modeling value $\mathbf{v}_k=(\text{SampEn}_k,|\rho_{i,k}|)$
%    }
%}

%\textbf{Value-oriented differentiation:} \\
%\For{each cluster $j=1:K$}
%{
%\Repeat{centroids $\psi_j$ converge}
%    {

%        Update centroid $\psi_j$ as the mean of all modeling values in cluster $j$ \\
%            \For{each attribute value $\mathbf{v}_k$}
%            {
%                Assign $\mathbf{v}_k$ to the nearest centroid
%            }
%    }
%}
%Differentiate attributes to L1 to L4 based on clusters
%\caption{Value-oriented attribute differentiation}\label{NAD-Alg}
%\end{algorithm}
In addition to the modeling cost determined by data complexity, it is equally important to analyze the modeling benefit that each network attribute can bring. This benefit is primarily assessed based on the relevance of the attribute to the operational objective. As the scope of this study is to maximize QoS satisfaction, for the same series of data $\mathbf{x_{i,k}}$, its modeling benefit can be reflected by its correlation with the $n$th QoS situation $\mathbf{Q_{i}^n} = \{Q_{i}^{n,l}\,|\,l=1,...,X_N\}$. To reduce complexity in HDT, we first adopt the Pearson correlation coefficient to evaluate the modeling benefit, given by
\begin{equation}
\label{CorrelationCoefficient}
\rho_{i,k}^n = \frac{\sum_{l=1}^{X_N}(x_{i,k}^l - \bar{x}_{i,k})(Q_{i}^{n,l} - \bar{Q}_{i}^n)}{\sqrt{\sum_{l=1}^{X_N}(x_{i,k}^l - \bar{x}_{i,k})^2\sum_{l=1}^{X_N} (Q_{i}^{n,l} - \bar{Q}_{i}^n)^2}},
\end{equation}
where $\bar{x}_{i,k}$ and $\bar{Q}_{i}^n$ represent the mean values of each attribute and operational objective, respectively. A higher $|\rho_{i,k}^n|$ indicates a stronger relationship between the $k$th attribute and $n$th objective, suggesting a higher benefit of modeling this particular attribute in user $i$. 

It is worth noting that $\rho_{i,k}^n$ only provides a basic observation of the data relevance. With a small $|\rho_{i,k}^n|$ for the $k$th attribute, e.g., $|\rho_{i,k}^n|<0.3$, it is worthwhile to further validate its irrelevancy through causality inference \cite{CausalityInference}, which can be utilized to obtain a deeper insight into the causal relationship among multiple attributes. For instance, Granger causality can be estimated by comparing the coarse-grained digital twin models for $\mathbf{Q_{i}^n}$ after adding $\mathbf{x_{i,k}}$, i.e., $\hat{Q}_{i}^n(X_N+1) = f_1(\mathbf{Q_{i}^n})$ and $\check{Q}_{i}^n(X_N+1) = f_2(\mathbf{Q_{i}^n}, \mathbf{x_{i,k}})$, where $f_1$ and $f_2$ are the modeling of the two cases typically through auto-regression. Based on the predictive power of $f_1$ and $f_2$ to compare the modeling accuracy, the causality relationship between $\mathbf{Q_{i}^n}$ and $\mathbf{x_{i,k}}$ can be obtained to update the modeling benefit $|\rho^n_{i,k}|$. Finally, by considering all QoS $\mathcal{Q}$, the modeling benefit for each attribute can be derived as $\rho_{i,k} = \sum_{n\in\mathcal{Q}}\omega_{i}^n\rho^n_{i,k}$.

After obtaining the modeling cost and benefit for each attribute, we can differentiate them into distinct groups for inclusion or exclusion during digital twin modeling. For the $k$th attribute from user $i$, its modeling value $\mathbf{v}_{i,k}=(\text{SampEn}_{i,k},|\rho_{i,k}|)$ can be used during differentiation to ensure the formation of several distinct groups with similar modeling values. This can be achieved by various methods, e.g., the K-means clustering algorithm, where the objective function of differentiation will be given by
\begin{equation}
    \label{KMeansObjective}
\min_\mathcal{G}\quad \sum\nolimits_{j=1}^{K} \sum\nolimits_{\mathbf{v}_k \in \mathcal{G}_j} | \mathbf{v}_k - \psi_j |^2,
\end{equation}
which aims to optimally obtain $K$ clusters that contain similar values. In (\ref{KMeansObjective}),  $\psi_j$ is the $j$th centroid randomly selected from all values during initialization, while $\mathcal{G}_j$ is the collection of the values in cluster $j$, i.e., $\mathcal{G}_j=\{\mathbf{v}_k|\Vert \mathbf{v}_k-\psi_j\Vert \leq \Vert \mathbf{v}_k-\psi_i\Vert, \forall i < K\}$. After a few iterations, all attributes can be differentiated into a few groups based on their modeling value for adaptive attribute selection during digital twin construction.

\textbf{\textit{Remark:}} In this study, as an illustrative example, we differentiate all attributes into four distinct levels with different selection priorities according to their modeling values. Some specific scenarios are provided as follows:

\textbullet~\textbf{L1}: High-valued attributes, e.g., traffic load and user activity, which are usually long-term, predictable, and highly relevant to the application. These attributes are prioritized in the digital twin representation for fast problem identification and modeled in HDT. 

\textbullet~\textbf{L2}/\textbf{L3}: Medium-valued attributes with a lower value due to high modeling cost or low modeling benefit, like QoS conditions and mobility. The real-time data of these short-term or complex attributes will be directly used in HDT for immediate monitoring and analysis. Meanwhile, they will also be selectively modeled in LDT with more available computational capabilities. Typically, $\mathcal{K}_s = \{\boldsymbol{\text{L}_1} \cup \boldsymbol{\text{L}_2} \cup \boldsymbol{\text{L}_3}\}$ are the attributes adaptively selected according to the network situation for the entire hierarchical digital twin paradigm.

\textbullet~\textbf{L4}: Other irrelevant or unpredictable attributes such as random network outages will be excluded from modeling due to their low values during network orchestration. 

By differentiating the digital twin modeling and data collection strategy for different levels of attributes, the communication and computation burdens associated with network orchestration can be relieved.

\subsection{User Activity-Guided Network Analysis and Segmentation}

In large-scale HetNets, BSs are ubiquitously deployed to facilitate seamless connectivity, while the inherent spatial distribution of user activities usually results in a non-uniform traffic load distribution. For BS $j$, any remote users $i$ with $a_{i,j}(t)=0,\forall t\geq0$ will neither establish connectivity nor consume network resources from $j$, thereby rendering them non-essential entities for $j$. Based on this observation, the entire network can be segmented into non-overlapping sub-networks for rapid situation evaluation, where the users in different sub-networks remain insulated. Given the user-centric traffic distribution, \textbf{L1} attributes, e.g., historical user activity patterns and user-BS connectivity will be paramount to support network segmentation.

To analyze the user-BS dynamics, the traffic $\delta_{i,j}^T$ over a period $T$ can be quantified by
\begin{equation}
\label{dataTraffic}
   \delta_{i,j}^T = \sum\nolimits_{t=0}^Ta_{i,j}^tr_{i,j}^t,
\end{equation}
which leads to load analysis in illustrating the demand-supply discrepancy of BS $j$. A binary load indicator can be given by
\begin{equation}
\label{Eq:14}
    \mathcal{L}_j = \begin{cases}
        0, & \text{if} \quad \sum_{i\in\mathcal{N}}\delta_{i,j}^T<\zeta\mathcal{S}_j,\\
        1, & \text{otherwise}.
    \end{cases} 
\end{equation}
where $\mathcal{S}_j$ is the theoretical network capacity in BS $j$ while $\zeta$ is the threshold level for indicating high loads.

To achieve efficient segmentation, graph partitioning is adopted to divide the entire network graph based on user activity. More specifically, for the entire network $G$, let $U_l = \{u_1,...,u_s\}$ be the $l$th segment, which can be obtained by minimizing the ratio between the cut and weight, given by
\begin{equation}
\min\nolimits_{U_i} \quad\sum\nolimits_{i\in \mathcal{N}} 1-\frac{f_c(U_i,U_i)}{f_c(U_i,\mathcal{V})},
\end{equation}
where $f_c(\cdot)$ is the cut function within the graph, defined as
\begin{equation}
    f_c(U_a,U_b) = \sum\nolimits_{u\in U_a}\sum\nolimits_{v\in U_b}\mathcal{A}_{u,v}(\tau).
\end{equation}
The adjacency matrix $\mathcal{A}(\tau)$ is used to represent the weight for each edge in the graph $G$ at instant $\tau$, given by
\begin{equation}
    \mathcal{A}_{i,j}(\tau)=
\begin{cases}
0, & \text{if } i=j,\\
\mathcal{X}_{i,j}\underbrace{\sum\nolimits_{l}\widetilde{\phi_l\delta_{i,j}^T(\tau-l)}}_\text{user traffic modeling in HDT}, & \text{otherwise},
\end{cases}
\end{equation}
where $\boldsymbol{\mathcal{X}} \in \mathbb{R}_{M\times N}$ is the potential link matrix, with $\mathcal{X}_{i,j} = 1$ if user $i$ can connect with BS $j$ and 0 otherwise. For a special scenario where all users can be connected with all BSs in the selected areas, $\boldsymbol{\mathcal{X}}$ will be a matrix of ones denoted by $J_{M\times N}$. In user traffic modeling, $\widetilde{(\cdot)}$ denotes the normalization to ensure $0\leq\mathcal{A}_{i,j}(\tau)\leq 1$, while $\phi$ is the digital twin coefficients of user demand $\mathcal{D}_{i,j}(\tau):= \sum\nolimits_{l}\phi_l\delta_{i,j}^T(\tau-l)$ obtained in HDT for modeling data traffic $\delta_{i,j}^T$ given in (\ref{dataTraffic}).

\subsection{Demand-Supply Analysis for Target Area Identification}

By focusing on each sub-network after segmentation, situations of the entire network can be efficiently evaluated. To avoid unnecessary resource consumption and delayed response, real-time data of short-term attributes (\textbf{L2}/\textbf{L3}), such as instantaneous link quality, QoS satisfaction, and user association, are used to analyze the current network states.

Let $\mathcal{Z}$ be the \textit{target area} identified within each network segment, which indicates an area with $N_\mathcal{Z}$ users and $M_\mathcal{Z}$ BSs can be efficiently orchestrated. First, we use orchestration efficiency to identify users for inclusion, defined as the expected QoS improvement for all users after orchestration over the resource consumed for digital twin, given by
\begin{equation}
\max_{y_i} \quad\eta=\frac{\sum_{i\in \mathcal{N}}\sum_ {n\in\mathcal{Q}}y_i\omega_{i}^n(1-\xi_{i}^n)\mathcal{B}_{i}^n}{\sum_{i\in \mathcal{N}}\sum_{k\in\mathcal{K}_s}y_i\mathcal{C}_{i,k}}.{\label{optiMain4}}
\end{equation}
In (\ref{optiMain4}), the numerator denotes the overall expected improvements throughout $\mathcal{Z}$ with $y_i$ as a binary variable indicating the inclusion of user $i$. $\mathcal{B}_{i}^n = \max(0, 1 - S_{i}^n)$ is the expected orchestration benefit, while $s_{i}^n$ is the QoS satisfaction indicator of user $i$, given by
\begin{equation}
\label{Eq:19}
    \xi_{i}^n = \begin{cases}
        0, & \text{if } S_{i}^n < 1,\\
        1, & \text{otherwise},
    \end{cases} 
\end{equation}
where the QoS satisfaction condition $S_{i}^n$ is based on the instantaneous network situations that are directly used in HDT. Moreover, the denominator of (\ref{optiMain4}) represents the resource consumption for establishing the digital twins to support the orchestration, with $\mathcal{C}_{i,k}$ as the resource used for digital twin construction for the $k$th attribute of user $i$.

On the other hand, the selection of target areas should guarantee sufficient resources to satisfy the user demands. For BS $j$, based on the predicted user demand $\mathcal{D}_i$ and network capacity $\mathcal{S}_j$ from HDT, (\ref{optiMain4}) should be further constrained by
\begin{equation}
\sum\nolimits_{i\in \mathcal{Z}}y_i\sum\nolimits_{j\in \mathcal{Z}}\mathcal{D}_{i,j} \leq \sum\nolimits_{j\in \mathcal{Z}}y_j \mathcal{S}_j\sum\nolimits_{i\in \mathcal{Z}}\mathcal{X}_{i,j}/M_\mathcal{Z}N_\mathcal{Z},
\end{equation}
The scaling factor $\sum\nolimits_{i\in \mathcal{Z}}\mathcal{X}_{i,j}/M_\mathcal{Z}N_\mathcal{Z}$ aims to ensure sufficient resources within $\mathcal{Z}$ for each BS $j$.

Problem (\ref{optiMain4}) can be further converted into a graph partitioning problem after adjusting the adjacency matrix $\mathcal{A}_{i,j}$ according to current network situation $s_i$ and $\mathcal{L}_i$. Specifically, a higher weight will be assigned to unsatisfied users and lower-loaded BSs to ensure their selection during area identification. The weight matrix can be updated as
\begin{equation}
    \widehat{\mathcal{A}}_{i,j}(\tau) = (1-w_{i,j})\mathcal{A}_{i,j}(\tau-1) + w_{i,j}\Delta\mathcal{A}_{i,j}(\tau),
\end{equation}
where $w_{i,j}$ is the weighting factor, while $\Delta\boldsymbol{\mathcal{A}(\tau)}$ is used to update the adjacency matrix obtained based on real-time network situations evaluated by (\ref{Eq:14}) and (\ref{Eq:19}), written as
\begin{equation}
    \Delta\mathcal{A}_{i,j}(\tau) =
    \begin{cases}
        \max\Bigl(\frac{Q_{i}^n}{\widehat{Q}_{i}^n},1-\frac{\sum_{i\in\mathcal{N}}\delta_{i,j}^T}{\zeta\mathcal{S}_j}\Bigr), & \text{if } \xi_i\mathcal{L}_j = 0, \\
        0, & \text{otherwise}.
    \end{cases}
\end{equation}
The two graph partitioning problems can be straightforwardly solved by adopting spectral methods \cite{SpetralMethods}. To accommodate for the urgent QoS needs of critical users identified in (\ref{optiMain4}), other relevant BSs and users with a strong connection indicated by $\cal A$ should also be included, finalizing the target area $\mathcal{Z}$. Unselected areas with sufficient QoS conditions will remain the current orchestration policy with continuous monitoring for future target area identification.

\section{Network Optimization in Lower Layers Through Scalable Network Modeling}
\label{LDT-Section}

\begin{figure*}[t]
    \centering
    \includegraphics[width=15.5cm]{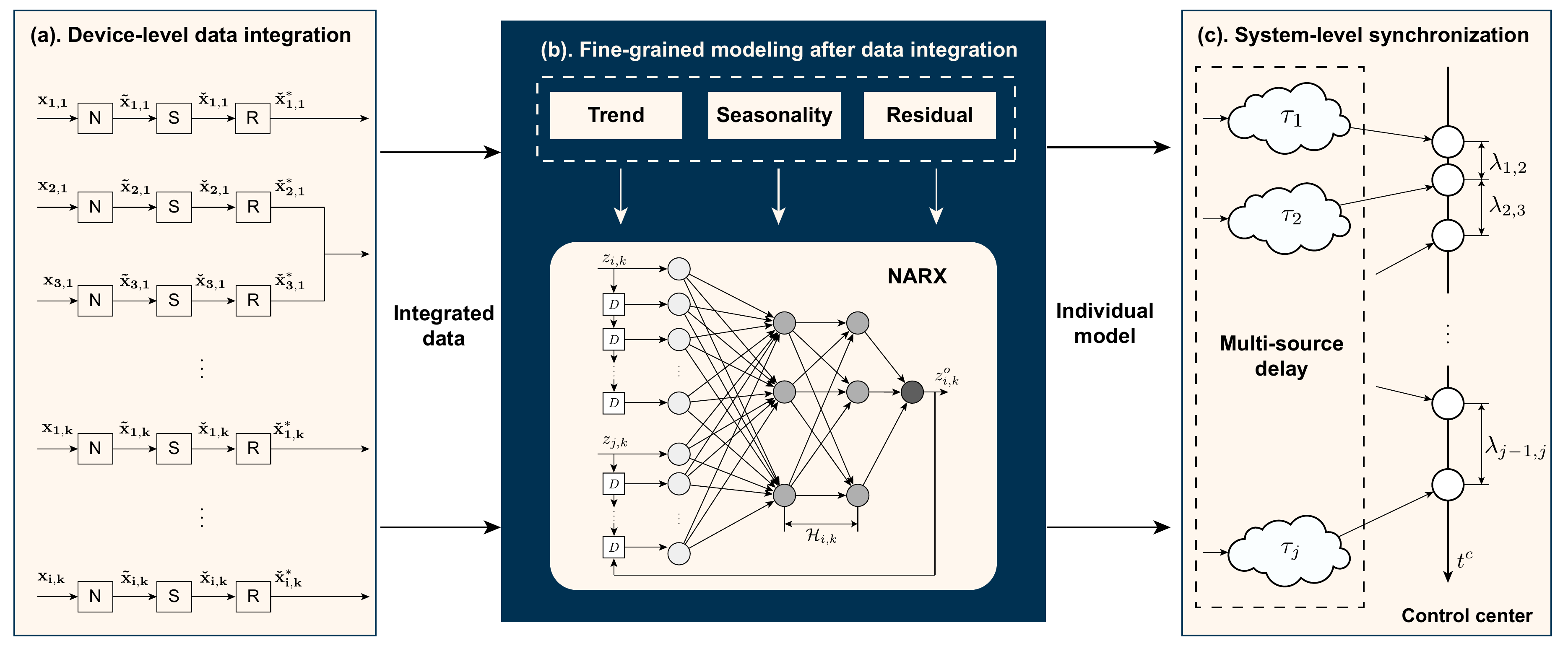}
    \caption{Illustration of the accurate digital twin construction in lower layers to support network orchestration: (a). Device-level data integration consists of data normalization, synchronization, and resampling to unify all data in the temporal domain. (b). Fine-grained modeling based on the integrated data using STL-NARX for individual or fused digital twins. (c). Cross-domain synchronization for all digital twins to address the model misalignment induced by multi-source modeling delays.}
    \label{fig:LDT}
\end{figure*}

In this section, we will focus on the scalable and fine-grained modeling of lower-layered digital twins to solve the network orchestration problem $\mathbf{\mathcal{P}4}$ for each target area $\mathcal{Z}$ identified from HDT. Moreover, due to the involvement of more attributes $\mathcal{K}_s$ in LDT modeling, the misalignment between physical and virtual domains becomes increasingly critical. Therefore, we designed a multi-level virtual-physical domain (VPD) synchronization mechanism to enhance the temporal consistency between virtual and physical domains across digital twin construction and utilization.

\subsection{Data Processing with Device-Level Synchronization}

Temporal discrepancies between the physical world and the digital domain are generally tolerable in HDT due to their primary focus on system-level target identification. In contrast, LDT necessitates more detailed modeling of the elements within the identified area $\mathcal{Z}$ to facilitate precise decision-making. The accuracy of the modeling process in LDT is affected by both pre-model data inaccuracies and post-model misalignment of the digital twin, adversely impacting network orchestration performance. In light of these challenges, we propose a multi-level synchronization strategy for LDT, as depicted in Fig. \ref{fig:LDT}, which addresses the issues of data disparity and model misalignment.

\subsubsection{Device-Level Data Integration}
Calculating awards $A_{i,j}$ in (\ref{Eq:4}) relies on accurate temporal alignment among $S_{i}^k$ for all users and their associated attributes. To maintain a unified understanding of each attribute in the control center, we streamline a three-step integration process to unify each data stream before individual model construction.

\textbullet~Data normalization: For the $k$th attribute of user $i$, its data $\mathbf{x}_{i,k}$ can be first scaled to a common range for enhanced comparability, given by
$\tilde{\mathbf{x}}_{i,k} = (\mathbf{x}_{i,k} - \mu_{\mathbf{x}_{i,k}})/(\sigma_{\mathbf{x}_{i,k}})$, where 
$\tilde{\mathbf{x}}_{i,k}$ denotes the normalized value, while $\mu_{\mathbf{x}_{i,k}}$ and $\sigma_{\mathbf{x}_{i,k}}$ are the mean and standard deviation of $\mathbf{x}_{i,k}$.

\textbullet~Data synchronization: Then, due to inherent variations in clock quality, the normalized data $\tilde{\mathbf{x}}_{i,k}$ is still misaligned in the temporal domain. Therefore, by selecting a series of $X_N$ data $\tilde{\mathbf{x}}_{i,k} = \{\tilde{x}_{i,k}^1,...,\tilde{x}_{i,k}^{X_N}\}$ sampled at a fixed interval $\beta_{i,k}$, their sampling timestamps $\mathbf{\mathcal{T}_{i,k}} = \{\mathcal{T}_{i,k}^1,...,\mathcal{T}_{i,k}^{X_N}\}$ can be used to estimated the local time error $e_{i,k}$. The error at a given instant $t$ compared to the control center is derived as \cite{sync4}
\begin{equation}\label{clockError}
  e_{i,k}(t) = \Biggl(\underbrace{\frac{\mathcal{T}_{i,k}^{X_N}-\mathcal{T}_{i,k}^1}{\beta_{i,k}({X_N}-1)}}_\text{clock skew} - 1\Biggr)t + \underbrace{\frac{t_{i,k}^R-t_{con}^R+t_{i,k}^T-t_{con}^T}{2}}_\text{clock offset}.
\end{equation}
The clock offset is estimated by exchanging two pairs of timestamped packets according to Precision Time Protocol, where $t^T$ and $t^R$ are the transmitting and receiving instants at the corresponding devices. Therefore, $e_{i,k}$ can be used to compensate each data sample in eliminating clock-induced data inaccuracy, i.e., $\check{x}_{i,k}(t)=\check{x}_{i,k}(t-e_{i,k}(t))$.

\textbullet~Data resampling: Furthermore, synchronized data could be still misaligned due to heterogeneous sampling capabilities, time-varying activity cycles, and random missing inputs, which becomes particularly critical for digital twin formation involving multi-attribute data fusion. To address this issue, synchronized data $\mathbf{\check{x}_{i,k}}$ can be resampled into a series of $\theta$ data at instants $\mathbf{\mathcal{T}_{i,k}^*}$, i.e., $\{\mathbf{\check{x}_{i,k}^*},\mathbf{\mathcal{T}_{i,k}^*}\} = \{\check{x}_{i,k}^{*,l},\mathcal{T}_{i,k}^{*,l}\}_{l=1}^{\theta}$, by adopting interpolation techniques such as Gaussian Process Regression (GPR) \cite{GPR}, which can recover the missing values and quantify the uncertainty associated with the resampled data. GPR relies on creating a kernel function $\boldsymbol{\mathcal{F}}$ to calculate covariance, e.g., $\boldsymbol{\mathcal{F}}_{(\mathbf{\mathcal{T}_{i,k}}, \mathbf{\mathcal{T}_{i,k}^{T}})}$ for the synchronized data. Then, the resampled data at given instants $\mathbf{\mathcal{T}_{i,k}^*}$ can be derived by
\begin{equation}
\label{resample}
    \mathbf{\check{x}_{i,k}^{*}} = \boldsymbol{\mathcal{F}^\mathbf{T}_{(\mathbf{\mathcal{T}_{i,k}}, \mathbf{\mathcal{T}_{i,k}^{*}})}}\bigl(\boldsymbol{\mathcal{F}}_{(\mathbf{\mathcal{T}_{i,k}}, \mathbf{\mathcal{T}_{i,k}^{T}})}+\sigma_f^2\mathbf{I}\bigr)^{-1}\mathbf{\check{x}_{i,k}},
\end{equation}
where $\sigma_f^2$ is the noise variance. The uncertainty of resampled data can be further evaluated by
\begin{equation}
\label{confidence}
    \boldsymbol{\Gamma_{i,k}^*}= \boldsymbol{\mathcal{F}_{(\mathbf{\mathcal{T}_{i,k}^{*}}, \mathbf{\mathcal{T}_{i,k}^{*}})}}- \mathbf{\check{x}_{i,k}^{*}}\mathbf{\check{x}_{i,k}}^{-1}\boldsymbol{\mathcal{F}_{(\mathbf{\mathcal{T}_{i,k}}, \mathbf{\mathcal{T}_{i,k}^{*}})}}.
\end{equation}
Therefore, $\mathbf{\check{x}_{i,k}^{*}}$ can be used in LDT to establish fine-grained models with dramatically enhanced temporal correlation.
%$\{\mathcal{T}_{i,k}, \check{x}_{i,k}\}_{i=1}^N$

%\subsection{Feature-level synchronization}

\subsection{Fine-Grained Digital Twin Modeling With STL-NARX}
\label{SectionIVB}

Directly modeling the temporally aligned dataset $\mathbf{\check{x}_{i,k}^{*}}$ could be highly challenging due to its highly complex nature incurred by the inherent nonlinearity, including long-term shifts in network demands and various short-term regular patterns for daily/weekly usage cycles. Motivated by the strong seasonality of network attributes, we propose the STL-NARX algorithm, which establishes digital twins by the nonlinear auto-regressive exogenous neural network (NARX) with seasonal and trend decomposition using Loess (STL), where STL can effectively understand and decompose the intricate temporal patterns in network attributes for accurate modeling.

More specifically, the integrated data $\mathbf{\check{x}_{i,k}^{*}}$ can be decomposed into trend $\mathbf{z_{i,k}^{tr}}$, seasonality $\mathbf{z_{i,k}^{sea}}$, and residual components $\mathbf{z_{i,k}^{res}}$, where $\mathbf{z_{i,k}^{res}}$ represents the irregularities and noise within the original data, given by
\begin{equation}
    \mathbf{z_{i,k}^{res}} = \mathbf{\check{x}_{i,k}^{*}} - \mathbf{z_{i,k}^{tr}} - \mathbf{z_{i,k}^{sea}},
\end{equation}
where $\mathbf{z_{i,k}^{tr}}$ and $\mathbf{z_{i,k}^{sea}}$ are obtained by a series of iterative processes, defined by
%$\mathbf{z_{i,k}^{sea}}(\kappa) = \text{Loess}\bigl(\mathbf{\check{x}_{i,k}^{*}} - \mathbf{z_{i,k}^{tr}}(\kappa-1)\bigr)$ and $\mathbf{z_{i,k}^{tr}}(\kappa) = \text{Loess}\bigl(\mathbf{\check{x}_{i,k}^{*}} - \mathbf{z_{i,k}^{sea}}(\kappa)\bigr)$,
\begin{equation}
\left\{
\begin{aligned}
    & \mathbf{z_{i,k}^{sea}}(\kappa) = \text{Loess}\bigl(\mathbf{\check{x}_{i,k}^{*}} - \mathbf{z_{i,k}^{tr}}(\kappa-1)\bigr), \\
    & \mathbf{z_{i,k}^{tr}}(\kappa) = \text{Loess}\bigl(\mathbf{\check{x}_{i,k}^{*}} - \mathbf{z_{i,k}^{sea}}(\kappa)\bigr),
\end{aligned}\right.
\end{equation}
where $\text{Loess}(\cdot)$ is the Loess operator for local smoothing, while the initial trend estimation $\mathbf{z_{i,k}^{tr}}(0)$ can be obtained by averaging the integrated data $\mathbf{\check{x}_{i,k}^{*}}$.

As shown in Fig. \ref{fig:LDT}(b), the refined terms $\mathbf{z_{i,k}^{tr}}$, $\mathbf{z_{i,k}^{sea}}$, and $\mathbf{z_{i,k}^{res}}$ can be respectively modeled. For clarity, we consider modeling $\mathbf{z_{i,k}^{res}}$ by using NARX as an example, which aims to capture the variation of residual component of $\mathbf{\check{x}_{i,k}^{*}}$ while considering potential exogenous inputs from other related attributes. The corresponding NARX model, denoted by a nonlinear mapping function $\mathcal{W}_{i,k}^{res}$, can be formulated as
\begin{equation}\label{NARX}
\hat{z}_{i,k}^{res,l} = \mathcal{W}_{i,k}^{res}\bigl(\mathbf{z_{i,k}^{res,R_i}}, \mathbf{z_{j,k}^{res,R_j}},\mathcal{H}_{i,k}^{res}\bigr) + \epsilon_l,
\end{equation}
where $\mathbf{z_{i,k}^{res,R_i}} = \{z_{i,k}^{res,l-u}\}_{u=1}^{R_i}$ is the internal input while $\mathbf{z_{j,k}^{res,l}} = \{z_{j,k}^{res,l-v}\}_{v=1}^{R_j}$ is one or more exogenous inputs to $\mathcal{W}_{i,k}^{res}$. The optimal performance of NARX training necessitates the careful tuning of the number of hidden layers, denoted as $\mathcal{H}_{i,k}^{res}$, alongside the maximum lags $R_i$ and $R_j$. Additionally, the error term at any given instant $l$ is represented by $\epsilon_l$. Consequently, through the integration of NARX models derived from (\ref{NARX}) for each decomposed component, STL-NARX can construct a more fine-grained digital twin for the $k$th attribute of user $i$, represented as $\mathbf{\hat{x}_{i,k}}=\mathbf{\hat{z}_{i,k}^{tr}} + \mathbf{\hat{z}_{i,k}^{sea}} + \mathbf{\hat{z}_{i,k}^{res}}$, which enables the prediction of its future behavior to enhance network orchestration efficiency.

\subsection{Digital Twin-Enabled Network Orchestration}
Based on the fine-grained digital twins $\mathbf{\hat{x}_{i,k}}$ for each attribute in all target areas $\mathcal{Z}$, network situations can be predicted to proactively orchestrate the HetNets in terms of user association and power allocation $\{\bf a, \bf p\}$. For the problem $\bf \mathcal{P}2$, optimized results at a given instant $t_u$ can be derived according to (\ref{Solution1}) as
\begin{equation}
    \{{\bf \hat{a}^{t_u}, \hat{p}^{t_u}}\} = \argmax_{\bf a, p} \frac{\sum\nolimits_{l}\sum\nolimits_{(i,j)\in\mathcal{Z}_l}a_{i,j}^{t_u}\widehat{A}_{i,j}^{t_u}}{\sum\nolimits_{l}{\widehat{\Theta}_{\mathcal{Z}_l}+\widehat{\Theta}_\text{HDT}}},
\end{equation}
where $\widehat{A}_{i,j}^{t_u} =  \sum_{k\in\mathcal{Q}}\omega_{i}^nS_{i}^{n,{t_u}}(\bf p^{t_u})$ is the predicted assignment award between user $i$ and BS $j$ at instant $t_u$ for all selected attributes, while $\widehat{\Theta}_\text{HDT} + \sum_{l}\widehat{\Theta}_{\mathcal{Z}_l}$ is the total resource consumption determined by HDT as a constant. $\widehat{\Theta}_\text{HDT} \ll \sum_{l}\widehat{\Theta}_{\mathcal{Z}_l}$ represents the resource consumed for HDT and LDT modeling, respectively. For the $l$th identified area, the LDT resource consumption can be estimated by $\widehat{\Theta}_{\mathcal{Z}_l} = \sum_{i\in\mathcal{Z}_l}\sum_{k\in\mathcal{K}_s}\nu_{i,k}\mathcal{C}_{i,k}$, where $\nu_{i,k}$ is a binary indicator to be $1$ if attribute $(i,k)\in\mathcal{Z}_l\cap\{\textbf{L1}, \textbf{L2}, \textbf{L3}\}$ and 0 otherwise.

To generate accurate orchestration decisions at $t_u$, it is essential to ensure all digital twins are coherently synchronized with their physical counterparts at this specific instant. However, the multi-source delay induced during data transmission, processing, and modeling will inevitably lead to model-specific temporal disparities, necessitating synchronization across virtual and physical domains when collaboratively supporting the network orchestration in the control center. 

Based on this observation, we design a VPD synchronization scheme to address the misalignment between physical and virtual domains. For the model $\mathbf{\hat{x}_{i,k}}$, depending on its causality relations, it can be established based on one or more series of integrated data. For each series of data, we use $t_{j}^\text{Last}$ to denote its last collecting instant. After adopting STL-NARX, the model can be obtained at instant $t_m$ expressed by $\mathcal{Y}_{i,k}^{t_m}:=\mathbf{\hat{x}_{i,k}^{t_m}}$, which, however, reflects the information about $t_{j}^\text{Last}$. To ensure a cohesive understanding of the digital twins, each model should be compensated for the multi-source delay before utilization, given by $\hat{\mathcal{Y}}_j^{t_m} = \mathcal{Y}_j^{t_m+\lambda_{j\rightarrow c}}$, where $\lambda_{j\rightarrow c} = t_u-t_j^\text{Last}$ represents the temporal disparity between data collection and model utilization at control center for each digital twin. Therefore, by leveraging these additional timestamps, we can ensure digital twins generated at $t_m$ will still be effective when they are utilized at $t_u$.

After VPD synchronization, the assignment problem can be solved, which is unbalanced given $N\neq M$. By manually inserting a series of virtual users $\hat{\mathcal{N}}$ and BSs $\hat{\mathcal{M}}$, a balanced assignment with $|\widetilde{\mathcal{M}}| = |\widetilde{\mathcal{N}}|$, where $\widetilde{\mathcal{N}} = \{\mathcal{N},\hat{\mathcal{N}}\}$ and $\widetilde{\mathcal{M}} = \{\mathcal{M},\hat{\mathcal{M}}\}$, can be derived, which can be efficiently addressed by various techniques, e.g., Hungarian algorithm with polynomial-time complexity. By solving the assignment problem, $\{{\bf \hat{a}, \hat{p}}\}$ can be proactively obtained to achieve optimized QoS satisfaction efficiency.

\section{Performance Evaluation}
Given that we solve the orchestration problem $\bf {\cal P}2$ sequentially in the HDT and LDT via decomposition, evaluating the performance of each layer independently is crucial. Moreover, in the simulation, we focus on thoroughly validating the aspects of precisely identifying target areas and the scalable modeling for fine-grained digital twins, aligning with the core strategies of the proposed hierarchical digital twin paradigm.
\begin{figure*}[t]
    \centering
    \includegraphics[width = 0.96\linewidth]{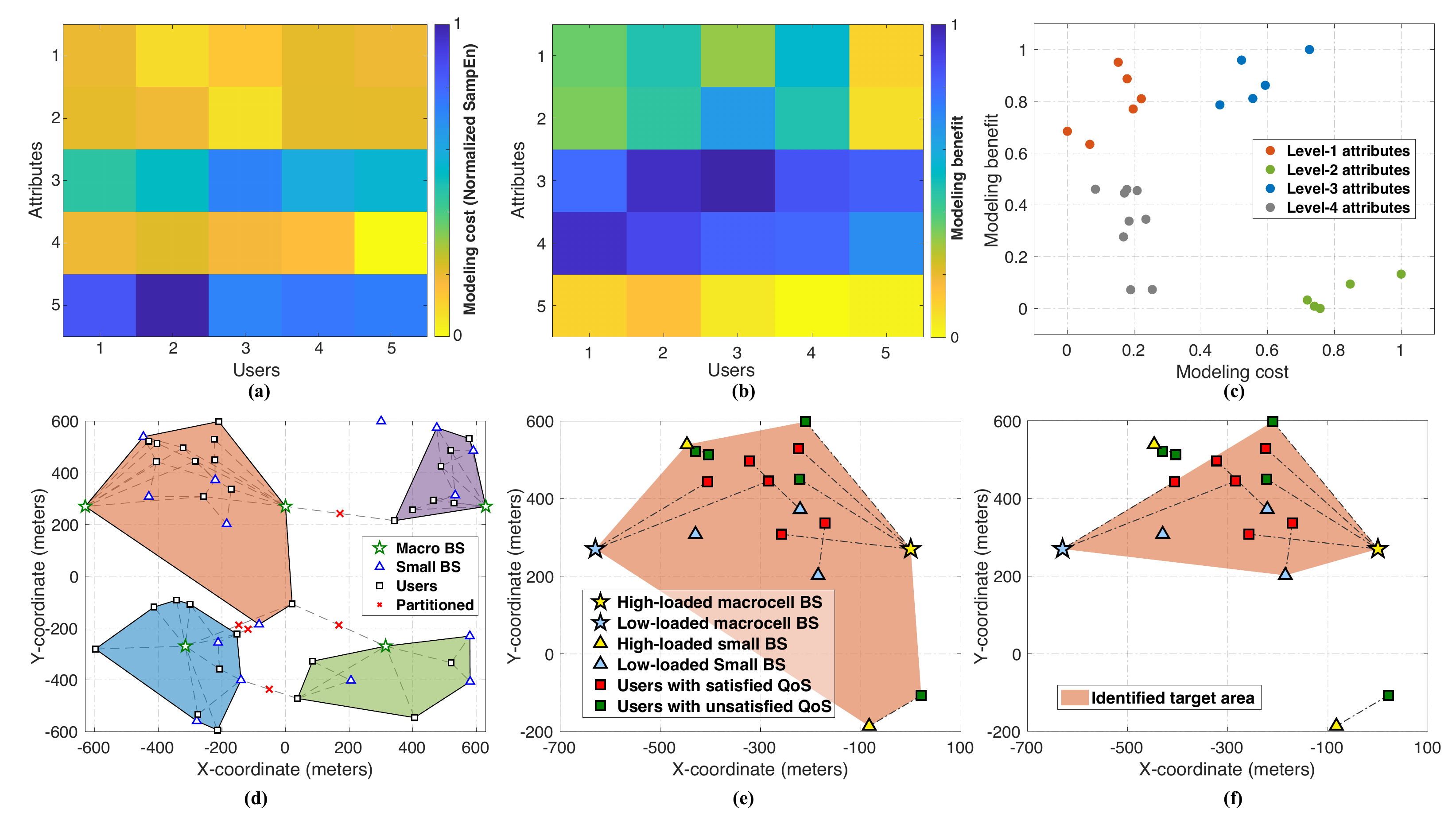}
    \caption{Identification of the target area in higher-layered digital twins based on differentiated network attributes and network situation analysis: (a). Modeling cost evaluated by sample entropy for different users and attributes. (b). Modeling benefit evaluated by correlation analysis and causality inference compared to the network objective. (c). Four-level attribute differentiation based on modeling value. (d). Segmentation of the HetNets into four sub-networks based on user activity. (e). Real-time network situation analysis of one sub-network, including traffic load of BSs and user QoS satisfaction. (f). Target area identification based on orchestration outcome for the selected sub-network.}
    \label{fig:SixFigures}
\end{figure*}

\subsection{Simulation setup}
\subsubsection{Environment of the HetNets}

The evaluation of our proposed hierarchical digital twin-enabled network orchestration scheme focuses on the HetNets comprising $5$ macro BSs with a maximum transmission power of $40$ dBm and a coverage radius of $300$ meters, uniformly distributed across the area. In addition, we opportunistically deploy $20$ small BSs within this network, each offering a maximum transmission power of $17$ dBm and covering a radius of $50$ meters. Distinct path loss models apply to macro cells $\bigl(15.3+ 37.6\text{log}_{10}(D)\bigr)$ and small cells $\bigl(8.46 + 20\text{log}_{10}(D)+0.7D\bigr)$, with $D$ denoting the distance between each user-BS pair. The total bandwidth of the system is $20$ MHz. Moreover, $200$ users are densely populated in certain areas to simulate environments like malls or offices. The channel fading model includes log-normal shadowing and small-scale fading, characterized by a background noise power of -104 dBm and a noise figure of 5 dB \cite{Fang}.
%\cite{Nange}

\subsubsection{Parameters for lower-layered digital twin modeling}
In the lower layers of our simulations, three distinct modeling techniques are employed, namely, auto-regressive integrated moving average (ARIMA), NARX, and our proposed STL-NARX approach. These models are crafted to predict a series of critical network attributes, including throughput, delay, traffic load, and packet loss within a specified area of the HetNets. Specifically, we configure the ARIMA models with parameters $(2,2,0)$ and fine-tune the NARX models using a learning rate of $0.001$, a delay setting of $10$, and $15$ hidden layers to enhance prediction robustness. Additionally, to consider daily and weekly network traffic patterns, seasonality periods are set to be $24$ and $168$ hours during STL analysis.

\subsection{Performance of higher-layered digital twins}
\subsubsection{Network attribute differentiation} As outlined in Section III-A, we assess the modeling value through a detailed cost-benefit analysis of various network elements, determining their impact on network performance. Figure \ref{fig:SixFigures}(a) presents the modeling costs based on sample entropy across a range of user attributes. Attributes such as network throughput (Attribute 1), delay (Attribute 2), and packet loss rate (Attribute 4) exhibit lower modeling costs, attributed to their stable statistical patterns, thereby simplifying prediction efforts. Conversely, dynamically fluctuated attributes, including network traffic load (Attribute 3) and user mobility (Attribute 5), show increased entropy, thus indicating the need for more computational resources for accurate modeling.

On the other hand, Fig. \ref{fig:SixFigures}(b) illustrates the modeling benefits derived from correlation analysis and causality inference relative to the network objectives. Specifically, modeling attributes critical to network orchestration, such as network traffic load, exhibit benefits close to one, indicating a substantial return on modeling efforts. In contrast, attributes like instantaneous channel quality and user mobility offer diminished value for long-term network orchestration. It is worth noting that the modeling value varies among users for the same attribute due to diverse QoS requirements, which underscores the need for customized attribute differentiation for individual users.

Based on the assessed modeling value, we categorize all attributes into four levels, as Fig. \ref{fig:SixFigures}(c) demonstrates. $\bf \text{L}1$ attributes, which yield high benefits for low costs, are modeled in HDT with high priority to identify target areas in the HetNets. Attributes classified as $\bf \text{L}2$ and $\bf \text{L}3$, which are still crucial for network orchestration, will be designated for more granular modeling in the LDT to support detailed and problem-centered network orchestration.
\begin{figure*}[t]
  \centering
  \begin{minipage}{0.49\textwidth}
    \includegraphics[width=\linewidth]{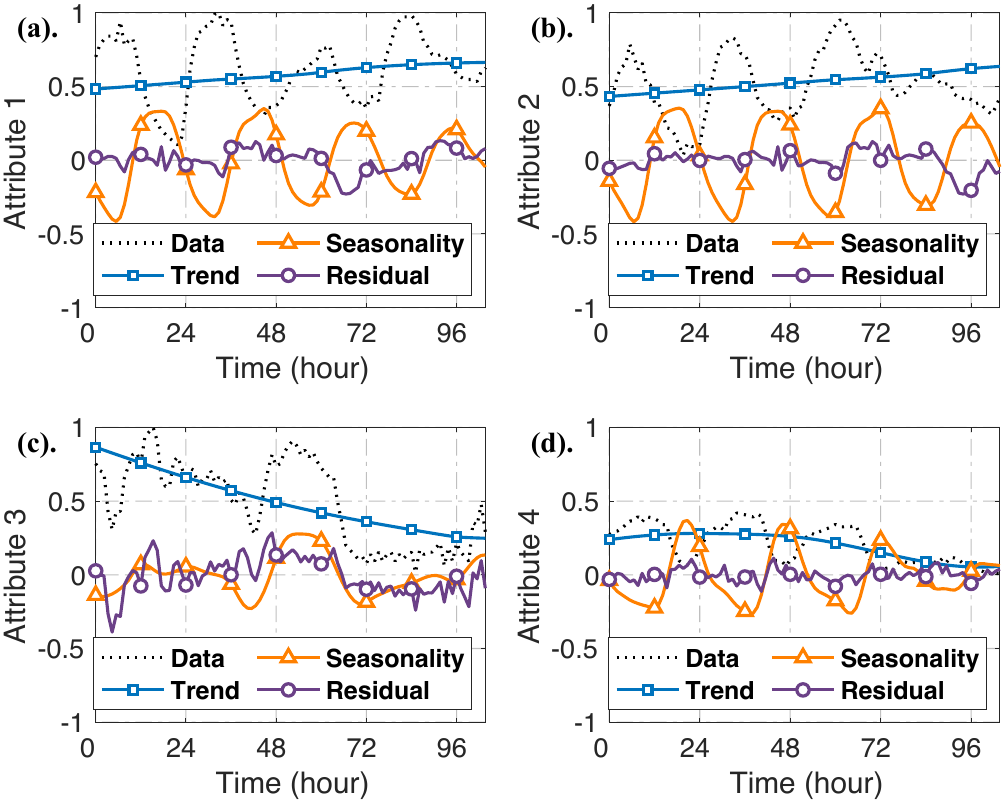}
    \caption{The analysis of the four attributes with STL, which decomposes data into trend, seasonality, and residual.}
    \label{Fig:decompSTL}
  \end{minipage}
\hfill
\hfill
  \begin{minipage}{0.48\textwidth}

    \includegraphics[width=\linewidth]{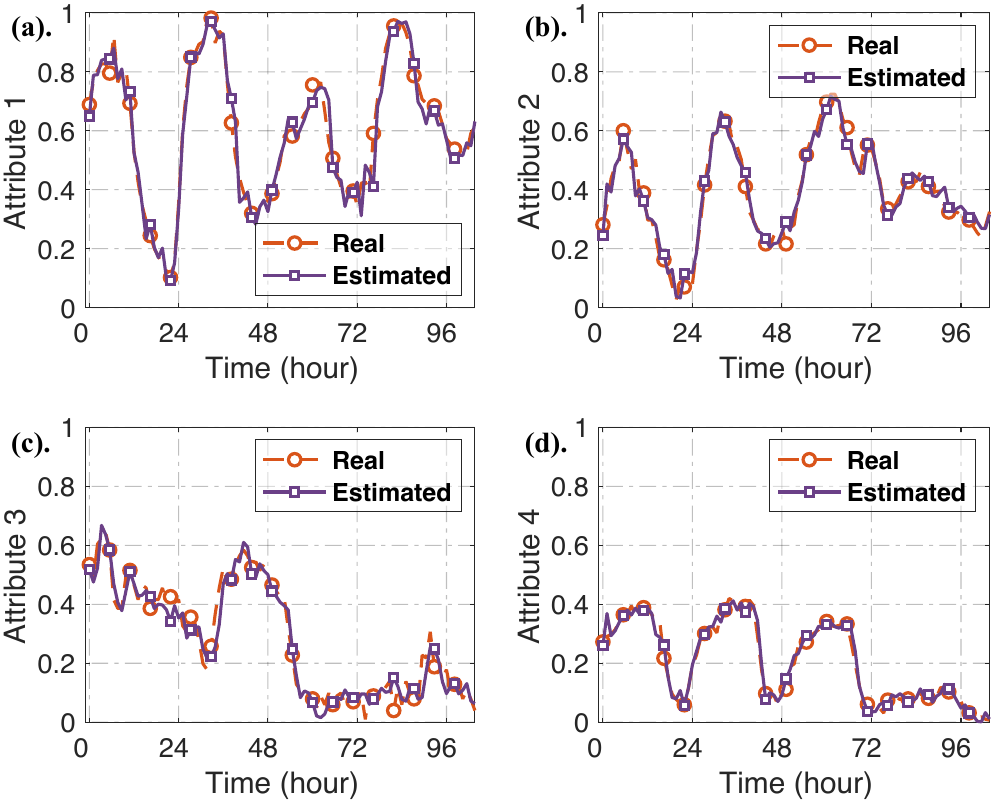}
    \caption{Accurate modeling is achieved by the proposed STL-NARX algorithm for the selected four attributes.}
    \label{Fig:Modeling}
  \end{minipage}
\end{figure*}

\begin{figure}
  \centering
  \includegraphics[width=.95\linewidth]{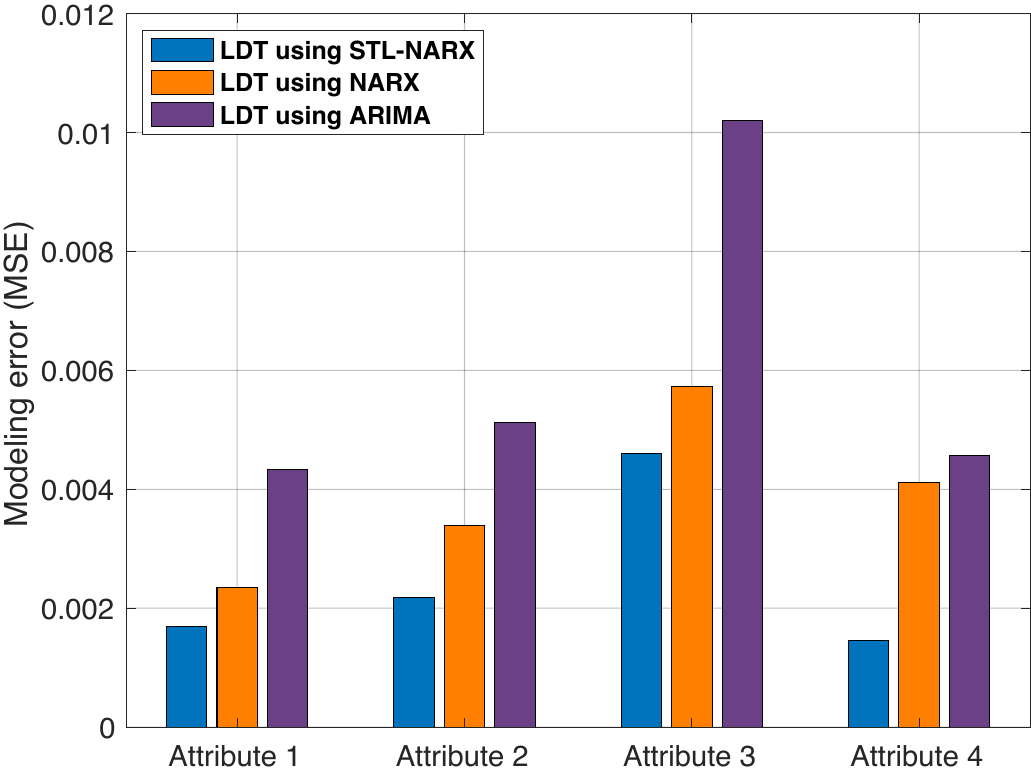}
  \caption{Comparison of three different modeling methods, including ARIMA, NARX, and STL-NARX, for the four selected attributes.}
  \label{Fig:modelingAcc}
\end{figure}

\begin{figure}
  \centering
  \includegraphics[width=.95\linewidth]{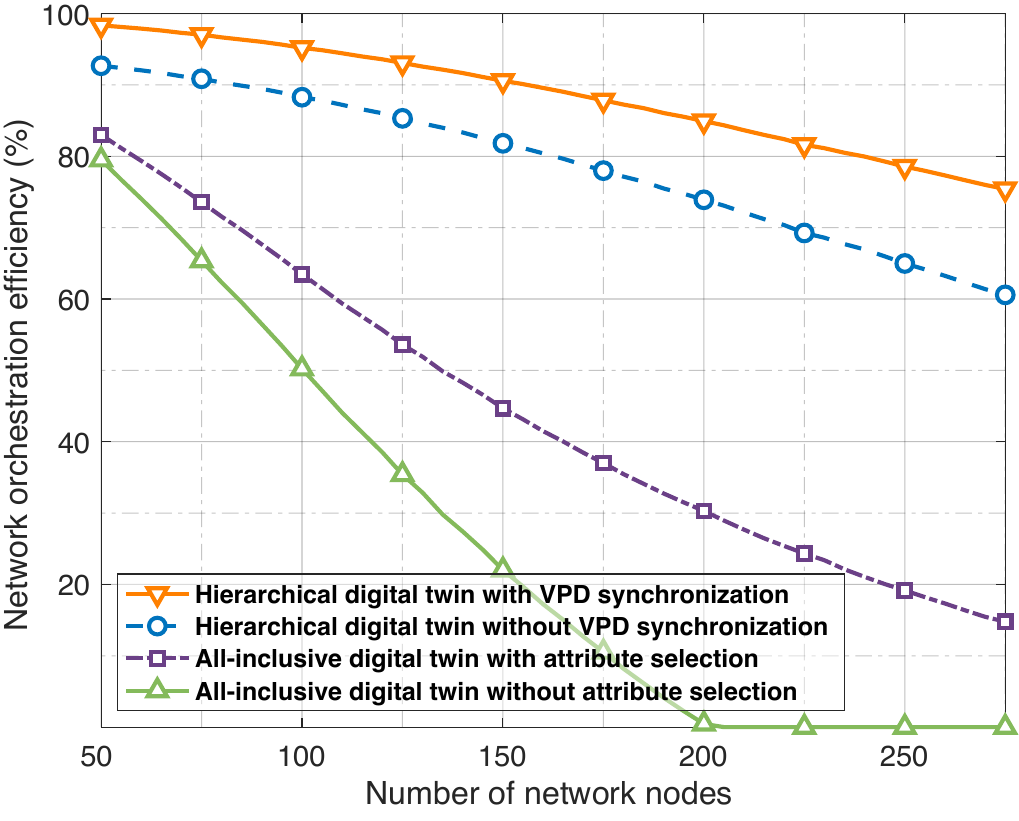}
  \caption{With different network scales, the hierarchical digital twin can always achieve the most efficient network orchestration.}
  \label{Fig:OrchEff}
\end{figure}
\subsubsection{Target area identification} Figure \ref{fig:SixFigures}(d) shows the layout of the HetNets, which is segmented into four discrete sub-networks using a graph partitioning algorithm. This division is based on the digital twins established in HDT using historical time-series data to identify areas with frequent user-BS interactions and high throughput, grouping them into cohesive sub-networks. In contrast, zones with sparse interactions are marked as segments with fragile connectivity, earmarked for selective disconnection due to their minimal impact on partitioning costs. Based on this network segmentation, a detailed examination of a specific sub-network is given in Fig. \ref{fig:SixFigures}(e), which shows the interaction between user density and BS capacity in fulfilling various QoS requirements. Clearly, regions with denser users are more likely to associate with lower QoS satisfaction and heavily burdened BSs, which should be prioritized during network orchestration in LDT.

Target areas within each sub-network are further identified by analyzing their expected cost-benefit for network orchestration, as shown in Fig. \ref{fig:SixFigures}(f). By adopting a refined graph partitioning process with adjusted cutting weights based on real-time network conditions, areas with the highest management effectiveness can be isolated. Therefore, QoS-satisfied users and high-loaded BSs, if not closely intertwined with other components, can be selectively filtered out. This ensures the prioritization of the most critical elements for detailed modeling in the construction of lower-layer digital twins.

\subsection{Performance of lower-layered digital twins}

According to the target areas identified, elements within the network that need particular management are selected for fine-grained modeling in LDT. Our STL-NARX approach first uses STL to decompose the time-series data generated from the selected users, where trend, seasonality, and residual components are isolated for precise analysis. As shown in Fig. \ref{Fig:decompSTL}, each series of data can be accurately decomposed into three components. Only daily seasonality is shown for clarity, where it clearly shows the daily activity pattern of throughput, delay, and packet loss rate regarding the selected user in Fig. \ref{Fig:decompSTL}(a), (b), and (d), respectively. Figure \ref{Fig:decompSTL}(c) illustrates that the network load for a nearby BS lacks a clear daily pattern, suggesting increased challenges in modeling accuracy. 

After decomposition, by focusing on each decomposed component, the predictability can be improved with better modeling performance. As shown in Fig. \ref{Fig:Modeling}, the proposed STL-NARX can accurately estimate each attribute over long periods. The dynamic nature of traffic load, influenced by other users as depicted in Fig. \ref{Fig:Modeling}(c), may lead to reduced modeling precision without incorporating exogenous inputs.

This can also be observed in Fig. \ref{Fig:modelingAcc}, where the training accuracy of three different techniques, i.e., ARIMA, NARX, and STL-NARX, are compared for the four attributes in terms of modeling mean squared error (MSE). With the integrated data, all three methods can obtain relatively accurate results, where the proposed STL-NARX outperforms the others, especially when pronounced seasonality is shown in the data. Attribute 3, which is more fluctuated over time due to external variations, can lead to higher modeling errors. However, by selecting relevant exogenous inputs from nearby users during digital twin modeling using NARX, the modeling error can be dramatically reduced compared to ARIMA. This highlights the significance of tight data integration and cross-domain synchronization for multi-attribute digital twin modeling.

The optimal network orchestration strategy is then concurrently achieved for each target area based on the prediction obtained from the digital twin models for efficient network QoS satisfaction. As shown in Fig. \ref{Fig:OrchEff}, the proposed hierarchical digital twin scheme achieves the highest efficiency for different network scales compared to the traditional all-inclusive frameworks, especially with synchronization across the two domains. Moreover, it can be observed that adaptive attribute selection and scalable network modeling can successively increase the network orchestration efficiency due to a better utilization of limited resources.

\section{Conclusion}
This paper has proposed a new hierarchical digital twin paradigm to achieve efficient network modeling to support 6G network orchestration through prioritized decomposition and selective modeling. By prioritizing highly valued network attributes within the higher-layered digital twin, rapid identification of the pressing network issues was facilitated. This allowed for the scalable modeling of fine-grained digital twins in the lower layers for all critical attributes and users in supporting efficient network optimization. In a nutshell, the proposed scheme can achieve situation-dependent problem identification in complex networks, leading to problem-centered solutions that significantly reduce resource wastage of traditional all-inclusive modeling approaches. Simulations have validated its capability to support efficient network orchestration in complex environments.
\label{Secion4}

\bibliographystyle{IEEEtran}
\bibliography{main}
\end{document}